\documentclass[superscriptaddress,nofootinbib,showkeys,twocolumn,prd]{revtex4}

\usepackage{amsmath}
\usepackage{amssymb}
\usepackage{amsfonts}
\usepackage{physics}
\usepackage{comment}
\usepackage{siunitx}
\usepackage{bm}

\newcommand{\beq}{\begin{equation}}
\newcommand{\eeq}{\end{equation}}
\newcommand{\ba}{\begin{eqnarray}}
\newcommand{\ea}{\end{eqnarray}}

\begin{document}

\title{On Coherent Delocalization in the Light-Matter Interaction}

\author{Nadine Stritzelberger}
\thanks{nadine.stritzelberger@cantab.net}
\affiliation{Department of Applied Mathematics, University of Waterloo, Waterloo, ON N2L 3G1, Canada}
\affiliation{Institute for Quantum Computing, University of Waterloo, Waterloo, ON N2L 3G1, Canada}
\affiliation{Perimeter Institute for Theoretical Physics,
Waterloo, ON N2L 2Y5, Canada}
\affiliation{Centre for Quantum Computation and Communication Technology, School of Mathematics and Physics, The University of Queensland, St. Lucia, QLD 4072, Australia}
\author{Achim Kempf}
\thanks{akempf@perimeterinstitute.ca}
\affiliation{Department of Applied Mathematics, University of Waterloo, Waterloo, ON N2L 3G1, Canada}
\affiliation{Institute for Quantum Computing, University of Waterloo, Waterloo, ON N2L 3G1, Canada}
\affiliation{Perimeter Institute for Theoretical Physics,
Waterloo, ON N2L 2Y5, Canada}
\affiliation{Centre for Quantum Computation and Communication Technology, School of Mathematics and Physics, The University of Queensland, St. Lucia, QLD 4072, Australia}
\affiliation{Department of Physics, University of Waterloo, Waterloo, ON N2L 3G1, Canada}

\begin{abstract}
We investigate how the coherent spreading of the center of mass wave function of a particle, such as an atom, molecule or ion, affects the particle's interaction with fields such as the electromagnetic field or a phonon field, in view also of possible applications to emerging quantum technologies. 
To this end, we develop a suitably generalized Unruh-deWitt model for the interaction between a delocalizing first quantized particle and a second quantized field. 
We study how the coherent spreading of the center of mass wave function of the particle affects emission and absorption rates and we find, in particular, that in the case of a supersonic coherent spreading in a medium, there should occur Cherenkov-like emissions, along with the excitation of the particle. 
\end{abstract}


\maketitle

\section{Introduction}

In the light-matter interaction, the motion of a particle, such as an atom, molecule or ion, influences the particle's emission and absorption properties through multiple effects. These effects range from the Doppler effect, that arises already in the non-relativistic regime, and the effect of Lorentz transformations on energy gaps at high velocities, to the Unruh effect that is expected to arise with extremely accelerated motion. In situations where the particle's  motion can be described by a classical probability distribution, these effects can be calculated separately for each possible state of motion, to then be added up incoherently. Our aim here is to investigate the case when the particle's motion is quantum uncertain. 

It is clear that there are differences between coherent and incoherent or classical superpositions of motion because quantum wave functions generically evolve differently than classical probability distributions. In fact, as we will discuss, differences can already occur between coherent and incoherent superpositions of the free evolution of the centre of mass of a particle, i.e., between coherent or incoherent free delocalization.  

In practice, to study emission and absorption processes for probability distributions of classical motion, it is usually convenient to transform into the various possible rest frames of the emitter or absorber, to then add up the effects incoherently. 
This strategy is, however, not straightforwardly applicable in the case of the coherent superpositions of quantized motion.


To avoid the need to transform into quantum uncertain rest frames, we will, therefore, employ a technical tool, previously used, e.g., in \cite{functional_calculus,maria_thesis}, that allows us to couple quantum fields to first-quantized particles that possess quantum uncertain positions. Technically, we will work with quantum fields that
take position operators as their argument, i.e., that are functions such as $\hat{\phi}(\hat{x})$ that are both operator dependent and operator valued. 
Further, we will work, for simplicity, in the non-relativistic regime and we will neglect all competing effects, such as higher ordere quantum field theoretic corrections. 

We will begin by modeling the light-matter interaction using a commonly employed idealization which focuses on only two energy levels of the matter system and which models the electromagnetic field as a scalar quantum field. When small matter systems, such as atoms, molecules or ions, are idealized as two-level qubit systems whose classical center of mass follows a prescribed trajectory, they are  known as Unruh-deWitt (UdW) detectors  \cite{Unruh_UdW_detector,DeWitt_UdW_detector}. These detectors have proven to be a very useful tool for the theoretical analysis of key processes such as the detection of Hawking and Unruh photons \cite{Unruh_UdW_detector,Gibbons_Hawking,Casadio_Venturi_1,Casadio_Venturi_2,birrell_davies} and, more recently, entanglement harvesting \cite{sorkin,Srednicki,Valentini_vacuum_entangled,Steeg_Menicucci_harvesting,Edu_Alejandro_harvesting1,Edu_Alejandro_harvesting2,Edu_Achim_Eric_William_farming} and quantum communication through quantum fields \cite{Cliche_Kempf_1,Cliche_Kempf_2,Jonsson_Martinez_Kempf}.

The conventional UdW detector model is, however, limited to the regime in which the center of mass follows a classical trajectory. 
We here generalize the UdW detector model to include the quantum mechanical description of its center of mass degrees of freedom. The dynamics of the center of mass wave function of the detector 
then effectively introduces an additional time dependence to the light-matter interaction. This additional time dependence arises already with the coherent spreading of the wave function under free time evolution. 
Our aim is to investigate the impact of this coherent delocalization on the light-matter interaction. 

We begin by showing that the spontaneous emission rate of an excited atom, molecule or ion can depend on the rate of its delocalization and on whether the delocalization is entirely coherent or in part also incoherent. We then show that a new phenomenon can arise in media, namely if parts of the center of mass wave function coherently spread faster than the maximum wave propagation speed in the medium. In this case, the coherent delocalization of the center of mass can trigger the excitation of the atom, molecule or ion, along with the emission of Cherenkov-like radiation. This leads to an effective friction and to decoherence for the supersonic contributions to the centre of mass wave function, possibly also leading to a Cherenkov-Zeno-like effect. These phenomena have the potential, for example, to impact the quantum channel capacities of light-matter interactions.

\section{The traditional UdW detector model for the light-matter interaction}

The traditional UdW detector model is a simplified model for light-matter interactions in which the electromagnetic field is modeled as a scalar massless quantum field. An atom, molecule or ion is then modelled as a first-quantized two-level system with ground state $\ket{g}$, excited state $\ket{e}$ and energy gap $E$.  The center of mass of a traditional UdW detector follows a prescribed classical worldline $\vec{x}(t)$, which we will here assume to be non-relativistic. 
The total Hilbert space of the coupled system factorizes,
$
\mathcal{H} = \mathcal{H}_{\text{internal}} \otimes \mathcal{H}_{\text{field}} $.
The interaction between the UdW detector and the quantum field is usually modeled as a linear coupling along the detector's worldline. 
In the Schr\"odinger picture, the interaction Hamiltonian 
takes the simple form:
\ba
\hat{H}_{int} &=& \lambda \hat{\mu} \otimes  \hat{\phi}(\vec{x}) \,\label{aaa}
\ea
In the literature, see, e.g., \cite{Edu_UdW_model}, the interaction Hamiltonian is sometimes extended to include a classical spatial smearing function to model the finite spatial extent of the detector's electronic orbits. We will here not make use of this technical tool. Instead, in Sec.\ref{section:hydrogen_atom}, we will describe the electronic orbitals explicitly. 
In Eq.(\ref{aaa}), $\lambda$ denotes the coupling strength,
$\hat{\mu}$ is the monopole operator of the detector, 
\ba
\hat{\mu} = \ket{e}\bra{g}+ \text{h.c.} \,,
\ea 
and $\hat{\phi}$ is the scalar quantum field,
\ba
\hat{\phi}(\vec{x})=\int \frac{d^3k}{(2\pi)^{3/2}} \sqrt{\frac{ c^2}{2k}} \, \Big[ e^{i\vec{k}\vec{x}}\hat{a}_{\vec{k}}+ \text{h.c.} \Big] \,.
\ea 
The coupling of a monopole to a scalar field  in Eq.(\ref{aaa}) is a simplified model for the coupling of a dipole to the electromagnetic field of the type $\hat{d}\cdot \hat{E}$. 
For a discussion of UdW-type interaction Hamiltonians, see, e.g., \cite{birrell_davies,Edu_UdW_model}. The free Hamiltonian of the UdW detector and the scalar quantum field is given by:
\ba
\hat{H}_0 =  E \ket{e}\bra{e} + \int \frac{d^3 k}{(2\pi)^{3/2}}  ck \, \hat{a}_{\vec{k}}^{\dagger}\hat{a}_{\vec{k}}^{\vphantom \dagger} \,
\ea
While $c$ here stands for the speed of light in the vacuum, we will later also consider media with lower wave propagation speeds. The transition probability for the system to evolve from an initial state $\ket{\Psi_{i}}$ at time $t_i$ to a final state $\ket{\Psi_{f}}$ at time $t_f$, working in the interaction picture and to first order perturbation theory, is obtained from the transition probability amplitude,
\ba
\label{eq:amplitude}
\mathcal{A} 
&=& -i \bra{\Psi_{f}} e^{-i\hat{H}_0 t_f} \int_{t_i}^{t_f} dt \, \hat{H}_{int}(t) \ket{\Psi_{i}} \,.
\ea
Here, $\hat{H}_{int}(t)$ denotes the interaction Hamiltonian in the interaction picture,
\ba
\hat{H}_{int}(t) &=& \lambda \hat{\mu}(t) \otimes \hat{\phi}(\vec{x},t) \,,
\ea
with $\hat{\mu}(t)$ and $\hat{\phi}(t)$ the monopole and field operators in the interaction picture,
\ba
\hat{\mu}(t) &=& e^{iEt}\ket{e}\bra{g} + \text{h.c.}  \,, \\
\hat{\phi}(\vec{x},t) &=& \int \frac{d^3k}{(2\pi)^{3/2}} \sqrt{\frac{ c^2}{2 k}} \Big[ e^{-ick t + i\vec{k}\vec{x}}\hat{a}_{\vec{k}}+ \text{h.c.} \Big]  \,.
\ea 
As an example, which we will later revisit, let us briefly review the spontaneous emission rate for an initially excited traditional UdW detector in the vacuum,
\ba
\ket{\Psi_{i}}=\ket{e}\otimes \ket{0}\,,
\ea
which is at rest, $\vec{x}(t)=\vec{x}_0$. We first consider the transition amplitude to a final state in which the detector is in its ground state and a field quantum of momentum $\vec{k}$ has been emitted,
\ba
\ket{\Psi_{f}}=\ket{g}\otimes \hat{a}^{\dagger}_{\vec{k}} \ket{0}\,.
\ea
We take the limits $t_i\rightarrow -\infty$ and $t_f\rightarrow \infty$ in order to eliminate switching effects. 
In order to avoid the divergence in the total spontaneous emission probability which arises from time translation invariance, see, e.g. \cite{birrell_davies}, we instead calculate the spontaneous emission rate, $\mathcal{R}_k$. Finally, to obtain the total spontaneous emission rate $\mathcal{R}$ irrespective of the momentum $\vec{k}$ of the emitted field quantum, we also trace over the Hilbert space of the field degrees of freedom. The calculation is straightforward and we here only state the result for later reference:  
\ba
\mathcal{R}
&=& \lambda^2E \, \label{eq:classical_decay_rate}
\ea
In the following section, our aim is to generalize the traditional UdW detector model in order  to take into account the effects that arise with the quantum delocalization of the detector. We will also allow the UdW detectors to couple to fields other than fundamental fields in the vacuum. For example, the UdW detector may couple to photons in dispersive media or to various fields of quasiparticles or collective excitations, such as spin waves or phonons in solids or in Bose Einstein condensates. This will allow us to consider scenarios where the UdW detector's real or virtual motion exceeds the speed of propagation of the quantum field that it couples to and we will find that new effects arise in this case.

\section{Generalizing the UdW detector model to include quantized center of mass degrees of freedom}

We will now go beyond the conventional model for UdW detectors, namely by dropping the simplifying assumption that the center of mass of the UdW detector follows a classical worldline. Instead, we will equip the UdW detector with first-quantized center of mass (CM) degrees of freedom. 
The total Hilbert space then factorizes as $
\mathcal{H} = \mathcal{H}_{\text{CM}} \otimes \mathcal{H}_{\text{internal}} \otimes \mathcal{H}_{\text{field}}
$.
We again model the interaction of the small quantum system and the quantum field via the monopole operator coupling. However, the coupling takes place at the center of mass position of the detector, which  is now described by the center of mass position operator $\hat{\vec{x}}$. That is, the interaction Hamiltonian becomes $
\hat{H}_{int} = \lambda \hat{\mu} \hat{\phi}(\hat{\vec{x}})
$. In order to make sense of the operator-valued field taking the position operator as its argument, we apply the spectral theorem, as described, e.g., in \cite{functional_calculus,maria_thesis}: an operator-valued function $\hat{f}$ can take an operator $\hat{A}$ as its argument by expanding the operator in its eigenbasis and evaluating the function on the operator's eigenvalues, $\hat{f}(\hat{A}) = \int da \, \ket{a}\bra{a} \otimes \hat{f}(a)$. Here, we obtain 
\ba
\hat{H}_{int} &=& \lambda \hat{\mu} \hat{\phi}(\hat{\vec{x}}) = \lambda \int d^3 x \ket{\vec{x}}\bra{\vec{x}} \otimes \hat{\mu} \otimes \hat{\phi}(\vec{x})\,, \quad
\ea
where $\ket{\vec{x}}$ are the position eigenstates and $\vec{x}$ are the position eigenvalues of the center of mass of the UdW detector.
The free Hamiltonians of the UdW detector and the scalar quantum field are given by
\ba
\hat{H}_0 = \frac{\hat{\vec{p}}^2}{2M} + E \ket{e}\bra{e} + \int \frac{d^3 k}{(2\pi)^{3/2}} c k \, \hat{a}_{\vec{k}}^{\dagger} \hat{a}_{\vec{k}}^{\vphantom \dagger} \,,
\ea
where $\hat{\vec{p}}$ denotes the center of mass momentum operator and $M$ is the mass of the UdW detector.
The interaction Hamiltonian, expressed in the interaction picture, then becomes
\ba
\hat{H}_{int}(t) &=& 
\lambda \int d^3 x  \ket{\vec{x}(t)}\bra{\vec{x}(t)} 
\otimes \hat{\mu}(t)
\otimes \hat{\phi}(\vec{x},t) \,, \quad
\ea
with the projection operators $\ket{\vec{x}(t)}\bra{\vec{x}(t)}$ evolving in the interaction picture according to:
\ba
\ket{\vec{x}(t)} &=& \int \frac{d^3 p}{(2\pi)^{3/2}}e^{-i\vec{p}\vec{x} + it\frac{\vec{p}^2}{2M}} \ket{\vec{p}} \,
\ea 
We are now ready to use Eq.(\ref{eq:amplitude}) to calculate transition probabilities for UdW detectors with coherently delocalizing center of mass.

\section{Spontaneous emission with quantum delocalizing center of mass}
\label{SE}

We begin by investigating the spontaneous emission rate of an UdW detector with quantized center of mass degrees of freedom, in order to then  compare the result to the spontaneous emission rate for a traditional UdW detector with classical center of mass.
Let us assume that the center of mass of the particle is prepared in an initial state $\ket{\varphi_0}=\int d^3 p \, \varphi_0(\vec{p}) \ket{\vec{p}}$.
The probability amplitude for the system to evolve from an initial state $\ket{\Psi_{i}} = \ket{\varphi_0} \otimes \ket{e} \otimes \ket{0}$ to a final state $\ket{\Psi_{f}} = \ket{\vec{p}\prime} \otimes \ket{g} \otimes \hat{a}^{\dagger}_{\vec{k}} \ket{0}$ becomes:
\ba
\mathcal{A}
&=& -  \frac{i\lambda}{\sqrt{2ck}} \frac{1}{(2\pi)^{9/2}}
e^{-i t_f \left(\frac{\vec{p}\prime^2}{2M} +ck \right)}
\int d^3 p \,\varphi_0(\vec{p})
\nonumber
\\
&&\times \int d^3 x\, e^{-i(\vec{p}\prime - \vec{p} + \vec{k})\vec{x}} 
\int_{t_i}^{t_f} dt \, e^{it\left(\frac{\vec{p}\prime^2-\vec{p}^2}{2M} - E + ck \right)} 
\nonumber
\\
&& + \mathcal{O}(\lambda^2)\,
\ea
Momentum conservation is automatically enforced, i.e., the momentum of the emitted photon and the recoil momentum of the detector are equal to the initial momentum of the detector. Energy is conserved as well, provided\footnote[3]{Finite $t_i$ and $t_f$ would correspond to a sudden on and off switching of the interaction by an external agent. As a consequence, time translation invariance would be broken and energy would not be conserved, since the agent could provide or extract energy to or from the system.} that we take the limits $t_i\rightarrow -\infty$ and $t_f\rightarrow \infty$. 
In order to obtain the total spontaneous emission rate $\mathcal{R}$ irrespective of the momentum $\vec{k}$ of the emitted photon or the recoil momentum $\vec{p}\prime$ of the detector, we  trace over the final state of the field and the external degrees of freedom of the particle:
\ba
\mathcal{R} &=& 
\frac{\lambda^2 c^2 M}{2} \int d^3 p |\varphi_0(\vec{p})|^2  \mathcal{T}(p) \,
\label{eq:decay_rate_general}
\ea 
Here, we defined 
\ba
\mathcal{T}(p) &:=& 2 
- \frac{1}{p}\sqrt{\left( p + Mc \right)^2 + 2EM}\nonumber  \\
&& + \frac{1}{p}\sqrt{\left( p - Mc \right)^2 + 2EM} \,,
\ea
with $p:=|\vec{p}|$. 
Since $\mathcal{T}$ does not depend on the initial center of mass wave function, we may call it the template function for the spontaneous emission rate. 

Let us now assume that the center of mass momentum distribution does not have significant amplitudes for large momenta. This allows us to Taylor expand the template function $\mathcal{T}$ around $p=0$, to obtain:
\ba
\mathcal{R} &=& \lambda^2 c^2 M A \int d^3 p |\varphi_0(\vec{p})|^2 \Big[ 1 +  (p/p_0)^2 + {\cal{O}}\left((p/p_0)^4\right)\Big] \nonumber\\
& \,&
\label{eq:decay_rate_expanded}
\ea
Here, we define the constants
\ba
A &:=& 1-\left(1+\frac{2E}{M c^2}\right)^{-1/2} \,,\\
B &:=& \frac{E}{c^4 M^3}\left( 1+\frac{2E}{M c^2} \right)^{-5/2} \,,\\
p_0 &:=& \sqrt{A/B}\,.
\ea
As is easily verified, in the regime where the energy gap is small compared to the mass energy of the detector, $E\ll Mc^2$, we have that $p_0\approx Mc$, i.e., the expansion in Eq.(\ref{eq:decay_rate_expanded}) is then valid in the non-relativistic regime. 

For instance, let us consider an ion that is initially localized in a quadratic potential of an ion trap \cite{Zoller,Ion_traps_Monroe,Ion_traps_Mukherjee}. After switching the ion trap off, the center of mass wave function of the ion will coherently spread. If the ion was prepared in an energy eigenstate of the trapping potential, the initial center of mass wave functions would be a Hermite function in three spatial dimensions. For example, if the ground state wave function of the center of mass is a Gaussian wave packet of initial width $L$, centered around $\vec{x}=\vec{x}_0$,
\ba
\ket{\varphi_0} &=& \int d^3 x \, \varphi_0(\vec{x},\vec{x}_0) \ket{\vec{x}}\,, \\
\varphi_0(\vec{x},\vec{x}_0) &=& \left( \frac{2}{\pi L^2} \right)^{3/4} e^{-\frac{|\vec{x}-\vec{x}_0|^2}{L^2}} \,,
\label{Gaussian_wavepacket}
\ea
we obtain that the spontaneous emission rate depends on $L$ through:
\ba
\mathcal{R}
&=& \lambda^2 c^2 M A \Big[ 1 +3 (L_0/L)^2 + {\cal{O}}\left((L_0/L)^4\right) \Big] \,
\label{decay_rate_Gaussian_wavepacket}
\ea
This approximation is valid for all $L\gg L_0$, where $L_0:=p_0^{-1}$ is effectively the Compton wavelength of the detector. This result shows that the faster the delocalization process, i.e., the sharper the initial localization, the more the spontaneous emission rate is increased. If instead the ion was prepared, for example, in the first exited eigenstate of the trapping potential in each direction, described by the product of the first (i.e., linear) Hermite polynomials and the Gaussian,
\ba
\ket{\varphi_0} = \int d^3 x \, \frac{8}{L^3} x_1 x_2 x_3 \,\varphi_0(\vec{x},0) \ket{\vec{x}}\,,
\ea
then too the wave function possesses more momentum, therefore spreads faster, and the spontaneous emission rate is further increased:
\ba
\mathcal{R}
&=& \lambda^2 c^2 M A \Big[ 1 + 9 (L_0/L)^2 + {\cal{O}}\left((L_0/L)^4\right) \Big] \,
\label{decay_rate_Hermite_wavepacket}
\ea

\section{Recovering the traditional UdW model in the limit of large mass and correspondingly slow delocalization}

Intuitively, the dynamical coherent delocalization of matter affects processes such as spontaneous emission because it introduces an effective time-dependence into the light-matter interaction. This suggests that in the limit of large detector mass, when the center of mass wave function coherently delocalizes more and more slowly, the spontaneous emission rate of the UdW detector with classical center of mass could be recovered.  

To verify this intuition, let us calculate the spontaneous emission rate in the limit of large detector mass. We expand the template function $\mathcal{T}$ for large detector mass $M$, i.e., for $Mc^2\gg E$ and $Mc\gg p$, to obtain to lowest order:
\ba
\mathcal{T}_0 = \frac{2E}{M c^2}
\ea
Since the 
momentum probability distribution is normalized, the integral in Eq.(\ref{eq:decay_rate_general}) can be carried out in the limit of large detector mass, yielding the spontaneous emission rate
\ba
\mathcal{R}_0
&=&  \frac{\lambda^2 c^2 M}{2} \int d^3p \, |\varphi_0(\vec{p})|^2 \, \mathcal{T}_0
= \lambda^2E \,.
\ea
Comparing with  Eq.(\ref{eq:classical_decay_rate}), this means that for a detector with quantized center of mass whose momentum distribution is centered around zero, the spontaneous emission rate in the infinite mass limit indeed coincides with the spontaneous emission rate for a traditional UdW detector at rest.
We therefore confirmed the intuition that it is not the amount of delocalization of the center of mass, but rather the dynamics of its delocalization that affects the spontaneous emission rate. We further conclude that the traditional UdW detector model is a good approximation only in the limit of large detector masses.

\section{Incoherent versus coherent delocalization}

The delocalization process of the center of mass can be coherent or incoherent, depending on the purity of the initial state. So far we assumed the center of mass of the detector to be in a pure initial state $\ket{\varphi_{0}}$ and we explicitly calculated the spontaneous emission rate for a Gaussian wave packet state. However, the center of mass of the detector could also be in a superposition of several wave packet states. For instance, the center of mass could initially be in a coherent superposition, $\ket{\varphi_0} \sim \ket{\xi} + \alpha \ket{\chi}$ with a phase $\alpha \in \mathbb{C} $, of two Gaussian wave packets centered around $\vec{x}=\vec{x}_0$ and $\vec{x}=-\vec{x}_0$ respectively,
\ba
\ket{\xi} &=& \int d^3 x \, \varphi_0(\vec{x},\vec{x}_0) \ket{\vec{x}}\,,  \\
\ket{\chi}&=& \int d^3 x \, \varphi_0(\vec{x},-\vec{x}_0) \ket{\vec{x}} \,,
\label{pure_state}
\ea 
Alternatively, the center of mass could initially be in a superposition which is in part also incoherent. For instance, the center of mass could be initially in the mixed state $\rho_0 = \frac{1}{2}\left( \ket{\xi}\bra{\xi} + \ket{\chi}\bra{\chi} \right) \,.$
The light-matter interaction indeed distinguishes between coherent and incoherent delocalization: due to translation invariance, the spontaneous emission rate for the partly incoherent superposition is the same as the spontaneous emission rate for a single Gaussian wave packet, as given by Eq.(\ref{decay_rate_Gaussian_wavepacket}). For the coherent superposition, however, we intuitively expect that the spontaneous emission rate could be affected by the interference between the two wave packets, except of course in the limits $x_0\rightarrow 0$ and $x_0\rightarrow \infty$, with $x_0:=|\vec{x}_0|$, in which the overlap of the two wave packets in position space is trivial. Indeed, we find that this is the case: the spontaneous emission rate for the coherent superposition,
\ba
\mathcal{R}
&=& 
\lambda^2 c^2 M A \Big[ 1
+ 3\left( 1 - f(x_0,\alpha)\right)  (L_0/L)^2 \nonumber \\
&& + {\cal{O}}\left((L_0/L)^4\right) \Big]  \,,
\ea
now depends both on the separation $2x_0$ and on the phase $\alpha$ between the two interfering wave packets:
\ba
f(x_0,\alpha) &:=& \frac{4 x_0^2}{3 L^2} \frac{2\Re(\alpha) e^{-2 x_0^2/L^2}}{1+|\alpha|^2+2\Re(\alpha) e^{-2 x_0^2/L^2}} \,
\ea
We notice that the incoherent and coherent case match not only in the limits $x_0\rightarrow 0$ and $x_0\rightarrow \infty$, but also for a purely imaginary phase, $\Re(\alpha)=0$, and whenever the two superposed wave functions are orthogonal, since the spontaneous emission rate only depends on the modulus squared of the initial center of mass wave function. 

\section{Can the dynamics of delocalization trigger excitation?}

In this section, we investigate whether, in media,
the dynamics of the delocalization process of the center of mass wave function of an UdW detector in its ground state is able to trigger the excitation of the UdW detector, along with the emission of a field quantum.
Intuitively, the reason for why such a process might happen is that virtual motion in a medium, similar to real motion in a medium, could incur a Cherenkov-like effect.

First, let us recall that a charged classical particle traveling at a constant velocity through the Minkowski vacuum will not spontaneously emit field quanta, since the exact same physical situation is encountered in its rest frame where it is clear that there is no energy available to create field quanta. 
In a medium, however, boosts are nontrivial and it is known that a charged classical particle that travels on an inertial trajectory can emit quanta, namely if it travels at a velocity faster than the propagation speed of waves in the medium \cite{Cherenkov_firt_publication, Frank_Tamm, Cherenkov_later_publication}. 


Important for our purposes here is that also UdW detectors, such as atoms, molecules and ions necessarily carry a monopole or dipole (or higher multipole) charge
as they couple to the field.
This suggests to consider the possibility of a Cherenkov-like effect for UdW detectors. 

While the classical Cherenkov effect arises for classical charges coupled to classical fields, an UdW detector couples not merely to a classical field but to a field that is quantized. Further, the UdW detector model allows us to investigate the possible excitation of the quantized internal degree of freedom of the UdW detector along with the emission of Cherenkov radiation. 

But also, and we will here focus on this new question, we can ask whether merely \textit{virtual} motion, in particular, virtual motion due to the dynamical coherent delocalization of the quantized center of mass, can trigger the emission of field quanta along with the excitation of the UdW detector. The idea is that this Cherenkov-like effect could arise due to that part of the center of mass wave function which corresponds to coherent delocalization with velocities faster than the propagation speed of waves in the medium. 

To this end, let us consider an UdW detector in its ground state, with quantized center of mass, coupled to a quantum field in its ground state:
\ba
\ket{\Psi_{i}} = \ket{\varphi_0} \otimes \ket{g} \otimes \ket{0} \,
\ea
We calculate the transition probability to a state in which the detector is excited and a field quantum has been emitted,
\ba
\ket{\Psi_{f}} = \ket{\vec{p}\prime} \otimes \ket{e} \otimes \hat{a}^{\dagger}_{\vec{k}} \ket{0} \,.
\ea
Through a calculation similar to the derivation of the spontaneous emission rate we discussed before, we now obtain the excitation rate
\ba
\mathcal{R} &=& \frac{\lambda^2 c^2}{2} \int d^3 p \, |\varphi_0(\vec{p})|^2 \, \mathcal{T}(p) \,,
\ea
where we again obtained a template function: 
\begin{eqnarray}
\mathcal{T}(p) &:=& \int_0^{\infty} dk \int_{-1}^{1} dz \,  k \, \delta\left(-\frac{pkz}{M} + \frac{ k^2}{2M} + E + ck \right) \nonumber \\
&=& \frac{2M}{p} \sqrt{\left( p-cM\right)^2-2EM} \, \nonumber\\
&& \times \Theta(p-Mc -\sqrt{2EM}) \,
\end{eqnarray}
The Heaviside step function $\Theta$ in the template function implies that a finite transition probability arises exclusively from those parts of the initial center of mass momentum distribution for which $p\geq Mc + \sqrt{2EM}$.
This means that the dynamics of center of mass delocalization, i.e., virtual motion alone, can indeed trigger the excitation of the detector and the emission of a field quantum. The condition is that at least parts of the center of mass wave function must spread faster than the critical velocity $v_{crit}:=c+\sqrt{2E/M}$
set by the maximum propagation speed, $c$, of waves in the medium and also by the energy gap, $E$, of the detector. We notice that, depending on the size of the detector gap, the critical velocity can be significantly larger than the wave propagation speed $c$ in the medium. 

The case of a charge without an internal degree of freedom is obtained as the limiting case $E\rightarrow 0$. In this limiting case, the interaction Hamiltonian commutes with the then vanishing free Hamiltonian of the internal degree of freedom.

Regarding the terminology, we refer to the excitation and radiation induced by a superluminal, or supersonic, coherent spreading of the center of mass wave function as a \textit{Cherenkov-like} effect. 

Concretely, for instance for an atom coupling to the electromagnetic field in a medium, we expect that sufficiently superluminal virtual center of mass velocities (i.e., velocities satisfying $v\geq v_{crit}$) can lead to the excitation of the atom and the emission of a photon. In the same way, sufficiently supersonic center of mass virtual velocities of an atom in a Bose Einstein condensate should lead to the excitation of the atom and the emission of a phonon. 
For Bose Einstein condensates \cite{BEC_speed_of_sound} the sound propagation speed can be as low as \si{\milli\metre/\second}, i.e., atoms with virtual velocities above this speed can still be well within the non-relativistic regime that we are working in here.

Generally, the Cherenkov-like effect leads to dissipative friction for any coherent delocalization above the critical velocity $v_{crit}$, (reminiscent to the Greisen-Zatsepin-Kuzmin (GZK) limit for the real motion of cosmic ray protons \cite{Greisen, Zatsepin-Kuzmin}). The Cherenkov-like effect, therefore, also represents a source of decoherence for virtual motion above the critical velocity. In practical applications of quantum technologies, this could mean, for example, that if an atom or molecule in a medium is to receive quantum information by absorbing a photon or other field quantum entangled with an ancilla, then that transfer of entanglement, i.e., of quantum information, is vulnerable to decoherence. The vulnerability arises from the Cherenkov-like effect if the absorption process localizes the absorbing atom or molecule too strongly, namely if, after the absorption, the center of mass wave function contains significant components above the critical velocity $v_{crit}$. We notice that $v_{crit}$ can be manipulated externally in as far as the energy gap of the UdW detector can be manipulated externally, e.g., via the Zeeman or Stark effects.

\section{Harmonic hydrogen atom coupling to electromagnetic field}
\label{section:hydrogen_atom}

While the UdW detector model is a simplified model of the light-matter interaction that allows one to efficiently investigate aspects of emission and absorption processes qualitatively, let us now generalize one of our results above to a quantitative order-of-magnitude analysis.  
Namely, as we saw in section \ref{SE}, the UdW detector model indicates that the dynamics of the coherent delocalization of an atom's center of mass should impact the rate of spontaneous emission. In order to estimate the order of magnitude of the effect, it would not be reliable to continue to model the atom's internal degree of freedom as a simple qubit\footnote[4]{The conventional UdW detector model (with classical center of mass) is routinely extended to account for the finite size of the atom due to the electronic orbital wave functions by introducing spatial smearing functions \cite{Edu_UdW_model}. Here, for increased accuracy, we instead quantize all degrees of freedom.}
coupling to a scalar field. Instead, let us calculate the spontaneous emission rate for a  hydrogen atom coupled to the electromagnetic field with the center of mass of the hydrogen atom dynamically delocalizing.
We model the electron and the proton in the hydrogen atom fully quantum mechanically (with position operators $\hat{\vec{x}}_e$ and $\hat{\vec{x}}_p$ and momentum operators $\hat{\vec{p}}_e$ and $\hat{\vec{p}}_p$), which respectively interact with the electromagnetic field via minimal coupling. 

The only simplification that we will use, to make the calculation of the order of magnitudes estimate easier, is to replace the Coulomb potential by a harmonic potential that is tuned such that the energy gap, $\hbar \Omega$, between ground and first excited states match that of the Coulomb potential. In the temporal gauge, the Hamiltonian of this harmonic hydrogen atom is
\ba
\hat{H}  &=& \frac{\left(\hat{\vec{p}}_p-q_p\hat{\vec{A}}(\hat{\vec{x}}_p)\right)^2}{2m_p}  + \frac{\left(\hat{\vec{p}}_e+q_e\hat{\vec{A}}(\hat{\vec{x}}_e)\right)^2}{2m_e} \\
&& + \int \frac{d^3 k}{(2\pi)^{3/2}} c \hbar k \sum_{s=1}^{2} \hat{a}_{\vec{k}}^{s\dagger} \hat{a}_{\vec{k}}^s 
+ \frac{\mu \Omega^2}{2} \left(\hat{\vec{x}}_p-\hat{\vec{x}}_e\right)^2 \,,\nonumber
\ea
where the electromagnetic field operators,
\ba
\hat{\vec{A}}(\vec{x}) = \int \frac{d^3 k}{(2\pi)^{3/2}} \sqrt{\frac{\hbar}{2\epsilon_0 ck}} \sum_{s=1}^{2} \vec{\epsilon}_s(\vec{k})
\Big[ \hat{a}_{\vec{k}}^s e^{i\vec{k}\vec{x}} + \text{h.c.} \Big] \,,\quad
\ea
couple respectively to the position operators of the electron and the proton.
In order for the model to describe ions as well, for now we allow for different charges of electron and core. Later, we will set $q_e=q_p=1.6\cdot 10^{-19}C$ for the hydrogen atom.
The interaction Hamiltonian reads
\ba
\hat{H}_{int} &:=& 
\frac{q_e \hat{\vec{p}}_e \hat{\vec{A}}(\hat{\vec{x}}_e)}{2m_e} 
- \frac{q_p \hat{\vec{p}}_p\hat{\vec{A}}(\hat{\vec{x}}_p)}{2m_p} 
+ \text{h.c.} \,,
\ea
where here, in the dipole approximation, we neglected the diamagnetic $A^2$ terms, which are of second order in the fine structure constant.
We now introduce relative and center of mass position operators,
$\hat{\vec{x}}_{\text{rel}}:= \hat{\vec{x}}_e-\hat{\vec{x}}_p$ 
and 
$\hat{\vec{x}}_{\text{CM}} := \frac{m_e}{M}\hat{\vec{x}}_e+\frac{m_p}{M}\hat{\vec{x}}_p$,
as well as their conjugate momentum operators, $\hat{\vec{p}}_{\text{rel}}$ 
and
$\hat{\vec{p}}_{\text{CM}}$,
with $M$ the total mass and $\mu$ the reduced mass of the atom. The total Hilbert space factorizes as $
\mathcal{H} = \mathcal{H}_{\text{CM}} \otimes \mathcal{H}_{\text{rel}}\otimes \mathcal{H}_{\text{field}}$.
In the new coordinates, we obtain for the interaction Hamiltonian:
\ba
\label{eq:int_Hamilitonian_hydrogen}
\hat{H}_{int} &=& \int d^3 x \int d^3 y \, \hat{\vec{p}}_{\text{CM}} \ket{\vec{x}}\bra{\vec{x}} 
\otimes \ket{\vec{y}}\bra{\vec{y}} 
\nonumber \\
&&\otimes \bigg[  \frac{q_e}{2M} \hat{\vec{A}}\left(\vec{x}+\tfrac{\mu}{m_e}\vec{y}\right)  -  \frac{q_p}{2M} \hat{\vec{A}}\left(\vec{x}-\tfrac{\mu}{m_p}\vec{y}\right) \bigg] 
\nonumber\\
&&+ \int d^3 x \int d^3 y \ket{\vec{x}}\bra{\vec{x}}\otimes  \hat{\vec{p}}_{\text{rel}} \ket{\vec{y}}\bra{\vec{y}} 
\nonumber\\
&&\otimes \bigg[ \frac{q_e}{2m_e} \hat{\vec{A}}\left(\vec{x}+\tfrac{\mu}{m_e}\vec{y}\right)  +\frac{q_p}{2m_p} \hat{\vec{A}}\left(\vec{x}-\tfrac{\mu}{m_p}\vec{y}\right) \bigg] 
\nonumber\\
&& + \text{h.c.}\,
\ea
The free Hamiltonian of the atom and the electromagnetic field becomes,
\ba
\hat{H}_0 &=& \frac{\hat{\vec{p}}_{\text{CM}}^2}{2M}  +\frac{\hat{\vec{p}}_{\text{rel}}^2}{2\mu}  
+ \int \frac{d^3 k}{(2\pi)^{3/2}} c \hbar k \sum_{s=1}^{2}\hat{a}_{\vec{k}}^{s\dagger} \hat{a}_{\vec{k}}^s  
+ \frac{\mu \Omega^2}{2} \hat{\vec{x}}_{\text{rel}}^2 \nonumber\\
&&
\ea
and it allows us to express the interaction Hamiltonian in Eq.(\ref{eq:int_Hamilitonian_hydrogen}) in the interaction picture. 

Let us now calculate the spontaneous emission rate for an initially excited atom with quantized center of mass, coupled to the vacuum state of the electromagnetic field,
\ba 
\ket{\Psi_{i}} &=& \ket{\varphi_0} \otimes \ket{e} \otimes \ket{0} \,.
\ea
We assume that the three-dimensional harmonic oscillator is in either one of its three first excited states:
\ba
\ket{e}=\ket{n_1,n_2,n_3} \in \{ \ket{1,0,0}, \ket{0,1,0},\ket{0,0,1}\} \,
\ea
We first calculate the transition probability amplitude for the initial state to evolve to the final state $\ket{\Psi_{f}} = \ket{\vec{p\prime}} \otimes \ket{g} \otimes \hat{a}^{s\dagger}_{\vec{k}} \ket{0}$, in which the atom is in its ground state, $\ket{g}=\ket{0,0,0}$, and a photon of momentum $\vec{k}$ and spin $s$ has been emitted. We obtain, working in the interaction picture and to first order in perturbation theory:
\ba
\mathcal{A}
&=& -\frac{i}{\hbar} e^{-\frac{i}{\hbar}\left( \frac{\vec{p}\prime^2}{2M} + \frac{3}{2} \hbar \Omega  + \hbar c k  \right) t_f } \varphi_0(\vec{p}\prime + \vec{k}) \sqrt{\frac{\hbar}{2\epsilon_0 c k}} 
\nonumber\\
&& \times \vec{\epsilon}_s(\vec{k}) \frac{1}{(2\pi)^{3/2}} 
\int_{t_i}^{t_f} dt \,
e^{\frac{i}{\hbar}\left(\frac{-2\hbar \vec{p}\prime\vec{k}-\hbar^2\vec{k}^2}{2M} - \hbar \Omega + \hbar c k \right)t} 
\nonumber\\
&&\times \Bigg(\int d^3 y \,\vec{p}\prime \, \psi_g (\vec{y}) \psi_{e}(\vec{y}) \bigg[ \frac{q_e}{M}e^{-i\frac{\mu}{m_e}\vec{k}\vec{y}}-\frac{q_p}{M}e^{i\frac{\mu}{m_p}\vec{k}\vec{y}} \bigg]
\nonumber\\
&& + \int d^3 p \,\vec{p} \, \widetilde{\psi}_g(\vec{p}) \bigg[ \frac{q_e}{m_e} \widetilde{\psi}_{e}(\vec{p}+\tfrac{\mu\hbar}{m_e}\vec{k}) \nonumber\\
&& + \frac{q_p}{m_p} \widetilde{\psi}_{e}(\vec{p}-\tfrac{\mu\hbar}{m_p}\vec{k}) \bigg] \Bigg)
+ \mathcal{O}(q^2)
\ea
Here, $\psi_g(\vec{y})$, $\psi_e(\vec{y})$, $\widetilde{\psi}_g(\vec{p})$ and $\widetilde{\psi}_e(\vec{p})$ are the ground state and first excited state wavefunctions of the harmonic oscillator in the position and momentum representations respectively.
We now average over the three first excited states of the harmonic oscillator and trace over the recoil momentum $\vec{p}\prime$ of the center of mass, as well as over the momentum $\vec{k}$ and spin $s$ of the emitted photon, such as to obtain the spontaneous emission rate:
\ba
\mathcal{R}
&=&  
\frac{\mu M \Omega}{ 2 \epsilon_0 c\hbar}
\int d^3 p \left| \varphi_0(\vec{p})\right|^2 \mathcal{T}(p)\,
\ea 
Here, we defined the template function,
\ba
\mathcal{T}(p):= \int_{k_{-}}^{k_{+}} dk  \, \frac{F(k)^2}{p}  
 \Bigg[ 1 + \frac{k^2 p^2}{2\Omega^2\hbar^2 M^2}
- \frac{G(k)^2}{2\Omega^2\hbar^2}  \Bigg] \,,
\ea
with
\ba
k_{\pm} &:=&  \pm p - cM + \sqrt{\left(\pm p - cM\right)^2 + 2 \Omega\hbar M }  \,,
\\
F(k) &:=&  \frac{q_e}{m_e} e^{-\mu k^2/(4\Omega\hbar m_e^2)} + \frac{q_p}{m_p} e^{-\mu k^2/(4\Omega\hbar m_p^2)} \,,\quad
\\
G(k) &:=& \frac{k^2}{2M} + ck - \hbar\Omega \,.
\ea
We carry out the $k$ integration in the template function $\mathcal{T}$ and Taylor expand around $p=0$, to obtain:
\ba 
\mathcal{R} &=&  \frac{\mu \Omega C}{ 2 \epsilon_0 c}
\int d^3 p \left| \varphi_0(\vec{p})\right|^2 \Big[ 1 + (p/p_0)^2 + \mathcal{O}\left( (p/p_0)^4 \right) \Big] \nonumber\\
& \,& 
\label{rate_hydrogen}
\ea
Here, we defined the constants 
\ba
C &\approx& 1.66\cdot 10^{21} \si{\ampere^2 \second^3\kilogram^{-2} \meter^{-2}}
\,,\\
D &\approx& 2.96 \cdot 10^{64} \si{\ampere^2 \second^5 \kilogram^{-4}\meter^{-4}}
\,,\\
p_0 &:=& \sqrt{C/D} \approx 2.37 \cdot 10^{-22} \si{\kilogram} \, \si{\meter}/\si{\second} \,.
\ea 
We note that the momentum $p_0$ corresponds to a velocity $v_0 \approx 1.42 \cdot 10^5 \si{m}/\si{s}$. The expansion in E.(\ref{rate_hydrogen}) is, therefore, valid in the non-relativistic regime, namely for all center of mass wave functions which possess significant probability amplitudes only for velocities $v\ll v_0$.
Considering a Gaussian wave packet for the initial center of mass wave function, the spontaneous emission rate becomes a function of the initial width, $L$, of the Gaussian wave packet:
\ba 
\label{abc}
\mathcal{R} &=&  \frac{\mu \Omega C}{ 2 \epsilon_0 c}
\bigg[ 1+3 (L_0/L)^2 + {\cal{O}}\left( (L_0/L)^4\right) \bigg] 
\ea
Here, we defined $L_0:= \hbar/p_0 \approx 2.80\cdot 10^{-12}\si{\meter}$.
The lowest order term, $\frac{\mu \Omega C}{ 2 \epsilon_0 c} \approx 6.86 \cdot 10^8 \si{s^{-1}}$, which does not depend on the initial center of mass wave function, is indeed roughly the spontaneous emission rate of an excited hydrogen atom ($\mathcal{R}\approx 6.27\cdot 10^{8} \si{\second}^{-1}$, see, e.g., \cite{Bethe_Salpeter}). This indicates that our description of the hydrogen atom as an electron bound to a proton via a harmonic potential, rather than a Coulomb potential, is a reasonably good quantitative model for our purposes here, in the sense that it yields the right orders of magnitude for the spontaneous emission rate from the first excited states. 

Let us now assume that the center of mass of the hydrogen atom is initially coherently localized to some moderate extent, for example, at the scale of the size of the hydrogen atom, $L=5.29\cdot10^{-11} \si{\meter}$. From Eq.(\ref{abc}), we obtain that this should lead to an increase of the spontaneous emission rate (compared to the spontaneous emission rate obtained for a harmonic hydrogen atom with initially completely delocalized center of mass) of $0.84 \%$. It is reasonable to expect a similar-sized effect for the hydrogen atom with Coulomb potential. Let us also address the question of the validity of the non-relativistic approximation for the motion of the center of mass in this scenario. Our choice for $L$ above implies an uncertainty in position of $\Delta x \approx 3.74 \cdot 10^{-11}\si{\meter}$, which, via the uncertainty principle and given the mass of the hydrogen atom, corresponds to an uncertainty in velocity of $\Delta v \approx 5.31 \cdot 10^3 \si{\meter/\second}$, which is within the non-relativistic regime. 

\section{Conclusions and outlook}

The formalism of UdW detectors  provides a simplified model of the light-matter interaction in which atoms, molecules or ions are modeled as two-level first-quantized systems (or qubits) with a classical center of mass that possesses a prescribed trajectory. The UdW model has proven to be useful for qualitative studies of a wide range of important phenomena, from the Unruh and Hawking effects to entanglement harvesting and quantum communication through quantum fields. Here, we generalized the UdW detector model to include the quantumness of the center of mass degrees of freedom. 
 
First, we found that the dynamics of the coherent delocalization of the center of mass influences the emission and absorption processes in the vacuum. 
This suggests that it should be very interesting to generalize prior studies with UdW detectors to include the quantumness of the center of mass of the UdW detectors. 
For example, the ability of a pair of UdW detectors to extract entanglement from the vacuum is known to depend on the spatial extent of the detectors \cite{Steeg_Menicucci_harvesting,Edu_Alejandro_harvesting1,Edu_Alejandro_harvesting2,Edu_Achim_Eric_William_farming}. It will be interesting, therefore, to examine also how the  quantum dynamics of the center of mass position uncertainty of UdW detectors modulates their ability to extract entanglement from the vacuum. These studies into the entanglement of the vacuum state could then also relate to holography, see, e.g.,
 \cite{Reznik_Retzker_Silman,sorkin,Srednicki,Valentini_vacuum_entangled,Summers_Werner_vacuum_entangled_1, Summers_Werner_vacuum_entangled_2,Harlow_review_holography,Susskind_Lindesay_holography}.
 
Second, we found the phenomenon that, in media,  the coherent delocalization of an atom, molecule or ion can induce Cherenkov-like radiation, along with the excitation of the particle. The phenomenon should occur when the virtual motion of the center of mass possesses  probability amplitudes for velocities faster than $v_{crit}=c +\sqrt{2E/M}$, where $c$ is the maximum wave propagation speed of the quantum field in the medium and where $E$ and $M$ are the particle's energy gap and mass respectively. 

This new Cherenkov-like effect may be experimentally observable, e.g., for an atom or molecule of a different species in a Bose Einstein condensate, if the particle coherently delocalizes faster than the velocity $v_{crit}$ that arises from the propagation speed, $c$, of phonons in the BEC and the energy gap, $E$. The sound propagation speed can be as low as \si{\milli\metre/\second} for certain Bose Einstein condensates \cite{BEC_speed_of_sound}. For quantitatively accurate predictions,  our calculations should, of course, be refined by using realistic dispersion relations in media, such as BECs. 

Several interesting consequences arise from the fact that the part of an atom, molecule or ion's center of mass wave function that coherently spreads faster than the critical speed $v_{crit}$ is prone to triggering the emission of the Cherenkov-like radiation. One consequence is that a rapid spread of the particle's center of mass wavefunction can be hindered by the energy loss (somewhat akin to evaporative cooling) due to the emission of Cherenkov-like radiation. 

On the other hand, it should be interesting to explore if the new Cherenkov-like effect may also lead to a more subtle Cherenkov-Zeno type of phenomenon in which it is not the spread of the particle's position wave function but in which it is instead the spread of the particle's momentum wave function which is hindered: let us consider a scenario where the particle or UdW detector is exposed to an external potential that induces the coherent spreading of its momentum wave function. For example, the particle could temporarily be in an inverted harmonic oscillator potential (which is feasible for trapped ions or atoms, see, e.g., \cite{Zoller,Ion_traps_Monroe,Ion_traps_Mukherjee}). In this case, as the UdW detector's center of mass momentum wave function tries to spread into large momenta, the medium continually keeps `measuring' whether or not among the coherent superpositions of states of motion of the UdW detector there are speeds above the critical speed $v_{crit}$, namely through the new Cherenkov-like effect. As a consequence, in a Cherenkov-Zeno-like effect, the spreading of the momentum wave function into these high momenta should be slowed down. 

It may also be possible to gain more intuition and insights into the predicted Cherenkov-like effect by using the new methods of quantum reference frames. To see this, let us first consider the regular Cherenkov effect: while a charge traveling with uniform speed below the wave propagation speed, $c_m$, in a medium will not radiate, the charge will radiate in the form of a shock wave if its speed exceeds $c_m$. Indeed, in a medium we can consider formal Lorentz transformations with $c$ replaced by $c_m$. A charge with a worldline that is formally spacelike with respect to $c_m$ would, after a suitable formal Lorentz transformation, be an extended charge that couples to the field at a point in time - and as such be bound to radiate. This explanation requires performing a formal Lorentz transformation that is specific to the speed of the worldline of the particle. 

In our case here, however, the motion of the center of mass is quantum and possesses a range of potential velocities in coherent superposition. This means that to extend our explanation above for the Cherenkov effect here requires one to perform coordinate changes to quantum uncertain reference frames via quantum uncertain Lorentz transformations. A formalism of such quantum reference frames and related techniques are being developed, see, e.g., \cite{QRF_Aharonov,QRF_Aharonov_Susskind,QRF_Bartlett_Rudolph_Spekkens,QRF_Palmer_Girelli_Bartlett,QRF_Giacomini_Castro-Ruiz_Brukner,wood_zych} and it will be natural to try to apply them to the Cherenkov-like effect here, that arises from coherent time evolutions of the center of mass, including coherent delocalization. The formalism of quantum reference frames may also be useful for taking into account relativistic effects, since it should allow us, for example, to hold the energy gap fixed in the detector's rest frame, even when the rest frame is quantum uncertain.

Finally, it should be very interesting to investigate the r\^ole of the quantumness of the center of mass degree of freedom of UdW detectors  in the transmission of quantum information, i.e., of entanglement, in the light-matter interaction. The transfer of entanglement between traditional UdW detectors via quantum fields has been studied in the field of relativistic quantum information, see, e.g.,  \cite{Cliche_Kempf_1,Cliche_Kempf_2,Jonsson_Martinez_Kempf}. The conventional UdW detector model is too crude, however, to capture some essential features, such as the quantum dynamics of recoil. 

Let us consider, for example, the case of a photon which is initially entangled with an ancilla and which is then absorbed by an atom. By absorbing the photon, the atom acquires the entanglement with the ancilla. The question arises to what extent it is the atom's center of mass degrees of freedom, and to what extent it is the atom's internal degrees of freedom that become entangled with the ancilla upon the absorption of the photon. 

The answer will depend, on one hand, on 
the amount by which the photon was entangled with the ancilla via its polarization and via its orbital degrees of freedom respectively. On the other hand, given the r\^ole of the recoil, the fraction of entanglement acquired by the center of mass degree of freedom will depend on
the dynamics of the delocalization of the atom's centre of mass. It should be very interesting, therefore, to generalize our investigation here for the study of quantum channels that arise with the light-matter interaction in modern quantum technologies, such as in quantum communication and quantum computing.

\begin{acknowledgments}
AK and NS are grateful to V. Sudhir, A. Jamison and T. Ralph for valuable discussions and to T. Ralph for his kind hospitality at the University of Queensland where part of this work was carried out. 
AK acknowledges support through the Discovery Program of the National Science and Engineering Research Council of Canada (NSERC). NS acknowledges support through an Ontario Trillium Scholarship and a Mitacs Globalink Research Award. 
\end{acknowledgments}

\bibliographystyle{ieeetr}
\bibliography{references}

\providecommand{\noopsort}[1]{}\providecommand{\singleletter}[1]{#1}%
\begin{thebibliography}{10}

\bibitem{functional_calculus}
W.~G. Unruh and R.~M. Wald, ``What happens when an accelerating observer
  detects a {Rindler} particle,'' {\em Phys. Rev. D}, vol.~29, pp.~1047--1056,
  1984.

\bibitem{maria_thesis}
M.~E. Papageorgiou, ``What is a field, what is a particle?...what about
  algebras?,'' Master's thesis, University of Waterloo, 2019.

\bibitem{Unruh_UdW_detector}
W.~G. Unruh, ``Notes on black-hole evaporation,'' {\em Phys. Rev. D}, vol.~14,
  pp.~870--892, 1976.

\bibitem{DeWitt_UdW_detector}
B.~S. DeWitt, {\em \rm "Quantum gravity: The new synthesis"}.
\newblock \it General Relativity, an {Einstein} Centenary Survey (eds. S. W.
  Hawking and W. Israel)\rm, Cambridge University Press, 1979.

\bibitem{Gibbons_Hawking}
G.~W. Gibbons and S.~W. Hawking, ``Cosmological event horizons, thermodynamics,
  and particle creation,'' {\em Phys. Rev. D}, vol.~15, pp.~2738--2751, 1977.

\bibitem{Casadio_Venturi_1}
R.~Casadio and G.~Venturi, ``The accelerated observer and quantum effects,''
  {\em Physics Letters A}, vol.~199, no.~1, pp.~33 -- 39, 1995.

\bibitem{Casadio_Venturi_2}
R.~Casadio and G.~Venturi, ``The accelerated observer with back-reaction
  effects,'' {\em Physics Letters A}, vol.~252, no.~3, pp.~109 -- 114, 1999.

\bibitem{birrell_davies}
N.~D. Birrell and P.~C.~W. Davies, {\em Quantum Fields in Curved Space}.
\newblock Cambridge Monographs on Mathematical Physics, Cambridge University
  Press, 1982.

\bibitem{sorkin}
R.~D. Sorkin, ``{On the entropy of the vacuum outside a horizon},'' {\em Tenth
  International Conference on General Relativity and Gravitation, Contributed
  Papers}, vol.~2, pp.~734--736, 1983.

\bibitem{Srednicki}
M.~Srednicki, ``Entropy and area,'' {\em Phys. Rev. Lett.}, vol.~71,
  pp.~666--669, 1993.

\bibitem{Valentini_vacuum_entangled}
A.~Valentini, ``Non-local correlations in quantum electrodynamics,'' {\em
  Physics Letters A}, vol.~153, no.~6, pp.~321 -- 325, 1991.

\bibitem{Steeg_Menicucci_harvesting}
G.~V. Steeg and N.~C. Menicucci, ``Entangling power of an expanding universe,''
  {\em Phys. Rev. D}, vol.~79, p.~044027, 2009.

\bibitem{Edu_Alejandro_harvesting1}
A.~Pozas-Kerstjens and E.~Mart\'{\i}n-Mart\'{\i}nez, ``Harvesting correlations
  from the quantum vacuum,'' {\em Physical Review D}, vol.~92, 2015.

\bibitem{Edu_Alejandro_harvesting2}
A.~Pozas-Kerstjens and E.~Mart\'{\i}n-Mart\'{\i}nez, ``Entanglement harvesting
  from the electromagnetic vacuum with hydrogen-like atoms,'' {\em Physical
  Review D}, vol.~94, 2016.

\bibitem{Edu_Achim_Eric_William_farming}
E.~Mart\'{\i}n-Mart\'{\i}nez, E.~G. Brown, W.~Donnelly, and A.~Kempf,
  ``Sustainable entanglement production from a quantum field,'' {\em Phys. Rev.
  A}, vol.~88, p.~052310, 2013.

\bibitem{Cliche_Kempf_1}
M.~Cliche and A.~Kempf, ``Relativistic quantum channel of communication through
  field quanta,'' {\em Phys. Rev. A}, vol.~81, p.~012330, 2010.

\bibitem{Cliche_Kempf_2}
M.~Cliche and A.~Kempf, ``Vacuum entanglement enhancement by a weak
  gravitational field,'' {\em Phys. Rev. D}, vol.~83, p.~045019, 2011.

\bibitem{Jonsson_Martinez_Kempf}
R.~H. Jonsson, E.~Mart\'{\i}n-Mart\'{\i}nez, and A.~Kempf, ``Information
  transmission without energy exchange,'' {\em Phys. Rev. Lett.}, vol.~114,
  p.~110505, 2015.

\bibitem{Edu_UdW_model}
E.~Mart\'{\i}n-Mart\'{\i}nez, M.~Montero, and M.~del Rey, ``Wavepacket
  detection with the {Unruh-DeWitt} model,'' {\em Phys. Rev.}, vol.~D87, no.~6,
  p.~064038, 2013.

\bibitem{Zoller}
D.~Jaksch, J.~García-Ripoll, J.~Cirac, and P.~Zoller, ``Quantum computing with
  cold ions and atoms: Theory,'' {\em Lectures on Quantum Information},
  pp.~391--422, 2007.

\bibitem{Ion_traps_Monroe}
C.~Monroe and J.~Kim, ``Scaling the ion trap quantum processor,'' {\em
  Science}, vol.~339, no.~6124, pp.~1164--1169, 2013.

\bibitem{Ion_traps_Mukherjee}
M.~Mukherjee, T.~Dutta, N.~V. Horne, Y.~Lei, P.~Liu, J.~Phua, and D.~Yum,
  ``({Invited}) {Ion} trap quantum computing: {A} new computing regime,'' {\em
  ECS Transactions}, vol.~86, no.~7, pp.~97--107, 2018.

\bibitem{Cherenkov_firt_publication}
P.~A. Cherenkov, ``{Visible emission of clean liquids by action of $\gamma$
  radiation},'' {\em Compt. Rend. Acad. Sci. URSS}, vol.~8, p.~451, 1934.

\bibitem{Frank_Tamm}
I.~M. Frank and I.~E. Tamm, ``{Coherent visible radiation of fast electrons
  passing through matter},'' {\em Compt. Rend. Acad. Sci. URSS}, vol.~14,
  no.~3, p.~109, 1937.

\bibitem{Cherenkov_later_publication}
P.~A. Cherenkov, ``Visible radiation produced by electrons moving in a medium
  with velocities exceeding that of light,'' {\em Phys. Rev.}, vol.~52, p.~378,
  1937.

\bibitem{BEC_speed_of_sound}
M.~R. Andrews, D.~M. Kurn, H.-J. Miesner, D.~S. Durfee, C.~G. Townsend,
  S.~Inouye, and W.~Ketterle, ``Propagation of sound in a {Bose-Einstein}
  condensate,'' {\em Phys. Rev. Lett.}, vol.~79, pp.~553--556, 1997.

\bibitem{Greisen}
K.~Greisen, ``End to the cosmic-ray spectrum?,'' {\em Phys. Rev. Lett.},
  vol.~16, p.~748, 1966.

\bibitem{Zatsepin-Kuzmin}
G.~T. Zatsepin and V.~A. Kuzmin, ``Upper limit of the spectrum of cosmic
  rays,'' {\em JETP Lett.}, vol.~4, pp.~78--80, 1966.

\bibitem{Bethe_Salpeter}
H.~A. Bethe and E.~E. Salpeter, {\em Quantum Mechanics of One- and Two-Electron
  Systems}, pp.~88--436.
\newblock Springer Berlin Heidelberg, 1957.

\bibitem{Reznik_Retzker_Silman}
B.~Reznik, A.~Retzker, and J.~Silman, ``Violating {Bell}'s inequalities in
  vacuum,'' {\em Phys. Rev. A}, vol.~71, p.~042104, 2005.

\bibitem{Summers_Werner_vacuum_entangled_1}
S.~J. Summers and R.~Werner, ``The vacuum violates {Bell's} inequalities,''
  {\em Physics Letters A}, vol.~110, no.~5, pp.~257 -- 259, 1985.

\bibitem{Summers_Werner_vacuum_entangled_2}
S.~J. Summers and R.~Werner, ``Maximal violation of {Bell's} inequalities is
  generic in quantum field theory,'' {\em Communications in Mathematical
  Physics}, vol.~110, no.~2, pp.~247--259, 1987.

\bibitem{Harlow_review_holography}
D.~Harlow, ``Jerusalem lectures on black holes and quantum information,'' {\em
  Rev. Mod. Phys.}, vol.~88, p.~015002, 2016.

\bibitem{Susskind_Lindesay_holography}
L.~Susskind and J.~Lindesay, {\em An Introduction to Black Holes, Information
  and the String Theory Revolution: The Holographic Universe}.
\newblock World Scientific, 2005.

\bibitem{QRF_Aharonov}
Y.~Aharonov and T.~Kaufherr, ``Quantum frames of reference,'' {\em Phys. Rev.
  D}, vol.~30, pp.~368--385, 1984.

\bibitem{QRF_Aharonov_Susskind}
Y.~Aharonov and L.~Susskind, ``Charge superselection rule,'' {\em Phys. Rev.},
  vol.~155, pp.~1428--1431, 1967.

\bibitem{QRF_Bartlett_Rudolph_Spekkens}
S.~D. Bartlett, T.~Rudolph, and R.~W. Spekkens, ``Reference frames,
  superselection rules, and quantum information,'' {\em Rev. Mod. Phys.},
  vol.~79, pp.~555--609, 2007.

\bibitem{QRF_Palmer_Girelli_Bartlett}
M.~C. Palmer, F.~Girelli, and S.~D. Bartlett, ``Changing quantum reference
  frames,'' {\em Phys. Rev. A}, vol.~89, p.~052121, 2014.

\bibitem{QRF_Giacomini_Castro-Ruiz_Brukner}
F.~Giacomini, E.~Castro-Ruiz, and {\v C}.~Brukner, ``Quantum mechanics and the
  covariance of physical laws in quantum reference frames,'' {\em Nature
  Communications}, vol.~10, 2019.

\bibitem{wood_zych}
C.~E. Wood and M.~Zych, ``Minimum uncertainty states for free particles with
  quantized mass-energy,'' {\em \rm arXiv:1911.06653}, 2019.

\end{thebibliography}

\end{document}